\documentstyle[osa,twocolumn,prl,aps,graphicx]{revtex}
\begin{document}
\wideabs{
\title{Vortices with antiferromagnetic cores in the SO(5) model of high temperature superconductivity}

\author{S. Alama$^1$, A.J. Berlinsky$^2$, L. Bronsard$^1$, and T. Giorgi$^1$}

\address{$^1$Department of Mathematics and Statistics,
McMaster University, \\
Hamilton, Ontario, Canada L8S 4M1}

\address{$^2$Department of Physics and Astronomy,
McMaster University, \\
Hamilton, Ontario, Canada L8S 4M1}

\date{\today}

\maketitle

\begin{abstract}
We consider the problem of superconducting Ginzburg-Landau (G-L) vortices with antiferromagnetic cores which arise in Zhang's SO(5) model of antiferromagnetism (AF) and high temperature superconductivity (SC).  This problem was previously considered by Arovas {\it et al.} who constructed approximate ``variational'' solutions, in the large $\kappa $ limit, to estimate the domain of stability of such vortices in the temperature-chemical potential phase diagram. By solving the G-L equations numerically for general $\kappa$,  we show that the amplitude of the antiferromagnetic component at the vortex core decreases to zero continuously at a critical value of the AF-SC anisotropy ($g\approx 0.25$) which is essentially independent of $\kappa$ for large $\kappa$. The magnetic field profile, the vortex line energy and the value of the B-field at the center of the vortex core, as functions of anisotropy are also presented.
\end{abstract}
}
%%%%%%% Whenever you want that R, type $\RR$ (or$\RR^d$ etc)
\def\RR{I\kern-0.35em R\kern0.2em \kern-0.2em}
%%%%%%% Whenever you want that N, type $\BBN$
\def\BBN{I\kern-0.35em  N\kern0.2em \kern-0.2em}

\newcommand{\be}{\begin{equation}}
\newcommand{\ee}{\end{equation}}
\newcommand{\bea}{\begin{eqnarray}}
\newcommand{\eea}{\end{eqnarray}}
\newcommand{\beann}{\begin{eqnarray*}}
\newcommand{\eeann}{\end{eqnarray*}}

\section{Introduction}

Zhang\cite{Zhang} has proposed a model which attempts to relate the antiferromagnetism and d-wave superconductivity observed in the high $\rm T_c$ cuprates by a symmetry principle.  In this model, the Neel vector, $\bf m$, which is the order parameter for antiferromagnetism, and the real and imaginary parts of the complex superconducting order parameter $\psi$ form the five components of a so-called ``superspin'' vector which transforms under the group SO(5).  This symmetry is explicitly broken by an anisotropy which favors antiferromagnetism (AF) for the undoped compounds (which have one unpaired electron spin per site). The anisotropy can be varied by a chemical potential term which adds holes (or equivalently removes electron spins).  For sufficiently large chemical potential, the anisotropy changes sign stabilizing the superconducting (SC) state. 

A straightforward consequence of this model, as noted by Zhang\cite{Zhang}, is that, close to the SC-AF phase boundary, vortices in the superconducting state should have antiferrimagnetic cores.  The idea is simply that, for small anisotropy, it is easier to simply rotate the order parameter in the core into the AF direction than it is to reduce its amplitude to zero. Arovas {\it et al.}\cite{Arovas}, following Zhang's suggestion, calculated the approximate domain of stability of vortices with AF cores, within Ginzburg-Landau theory, in the temperature-chemical potential plane. Their calculation compared the energies of two variational solutions for the behavior of the superspin vector, (1) the function $\tanh(r/\xi)$ for the magnitude of the SC order parameter, representing a vortex with a normal core and (2) a vector making an angle $\theta(r) = (\pi/2)exp(-r/\ell)$ toward the AF direction from the SC direction. The contribution of the vector potential to the free energy was!
 ignored, as is appropriate for 
the extreme type-II limit ($\kappa\rightarrow \infty$). By comparing the energies of these two configurations, minimized with respect to $\xi$ and $\ell$, as a function of the G-L parameters, Arovas and coworkers located a boundary within the SC region of the phase diagram, where vortex core solutions (1) and (2) exchange stability.  If these solutions were exact, the implication would be that the amplitude of antiferromagnetism in the core changes discontinuously across this boundary.

In this paper we consider the full numerical solution of the coupled non-linear Ginzburg-Landau equations for vortices in the phenomological SO(5) model, including the vector potential. We calculate the AF and SC order parameter profiles and the magnetic field profiles as functions of distance from the core for general $\kappa$ and anisotropy. We find a much broader range of stability for vortices with AF cores than was found by Arovas {\it et al.}  We also find that the amplitude of the AF component in the core falls continuously to zero with increasing chemical potential as the phase boundary for normal cores is approached, and we calculate the critical value of the anisotropy(chemical potential) where this transition occurs for a range of $\kappa$, including $\kappa=\infty$.  We also obtain the dependence of the vortex line energy and the field at the core on $\kappa$ in the large $\kappa$ limit for AF and normal cores.  Our results allow us to conclude that both $H_{c1}$ a!
nd $H_{c2}$ increase with increasing anisotropy as the cores become less antiferromagnetic.

\section{Free energy}
We consider the following Ginzburg-Landau free energy for superspin
vector ${\bf n}=(\psi_1,m_1,m_2,m_3,\psi_2)$ with vector potential ${\bf A}$,
\onecolumn
\twocolumn[

\be \label{1}
\frac12 \int \left\{ f_0+  a |{\bf n}|^2 + \frac12 b |{\bf n}|^4 - \tilde g |{\bf m}|^2
   +  {\hbar^2\over 2m^*} \left|\left({1\over i}\nabla-{e^*\over \hbar c}{\bf A}\right)\psi\right|^2 + 
         {\hbar^2\over 2m^*} |\nabla {\bf m}|^2 + {1\over 8\pi} |\nabla\times {\bf A}|^2
    \right\} \, d{\bf x}           
\ee

The anisotropy is measured by $\tilde g=4\chi (\mu_c^2-\mu^2)$,
where $\mu$ denotes the chemical potential.

We introduce a non-dimensional form of the energy.  In these units
the penetration depth $\lambda=1$, and the correlation length $\xi$
and GL parameter $\kappa$ are related via $\xi=\kappa^{-1}$.
More precisely, we let
\be  {\bf x}=\lambda \hat {\bf x},\qquad\psi=\sqrt{\frac{|a|}{b}}\hat\psi,  \qquad{\bf m}=
\sqrt{\frac{|a|}{b}}\hat{\bf m},\qquad
 {\bf A}=\frac{c\hbar}{\lambda e^*}\hat{\bf A},
\ee
where
\be \lambda=\sqrt{\frac{m^* b c^2}{4\pi |a|(e^*)^2}}\quad\hbox{ and }\quad
\xi=\sqrt{\frac{\hbar^2}{2m^*|a|}}.
\ee
Then the free energy becomes (up to a factor of $\frac{\hbar^2|a|}{2m^*b}$)
\be 
 \label{2}
 \frac12 \int \left\{  {\kappa^2\over 2}(1-  |\hat{\bf n}|^2)^2 
    + g\kappa^2 |\hat{\bf m}|^2   + \left|\left({1\over i}\nabla -\hat{\bf A}\right)\hat\psi\right|^2 + 
         |\nabla \hat{\bf m}|^2 + |\nabla\times \hat{\bf A}|^2
    \right\} \, d\hat{\bf x} .          
\ee
\vskip1pc
\indent
  \hfill\vrule depth1em height0pt \vrule width3.375in height.2pt depth.2pt
\vskip1pc
]

The anisotropy term is regulated by a new constant $g$,
\be\label{g}
     g= {\tilde g\over a} = {4\chi (\mu^2-\mu_c^2) \over |a|},
\ee
which is positive when superconductivity is preferred in the
bulk ($\mu>\mu_c$).
For convenience, we drop the hats on non-dimensional quantities in
the following.

\subsection{Vortex solutions}
We now seek vortex solutions in the plane, in other words free energy
minimizers of the form
$\psi(x)=f(r)e^{i\theta}$,  ${\bf m(x)}=m(r)\hat {\bf m}$
with $\hat {\bf m}$ a fixed unit vector, 
and 
${\bf A}(x)=[S(r)/r^2](-y,x)=[S(r)/r]\hat{\bf \theta}.$
In addition we require that $f(0)=0,$ $f(r\to\infty)\to 1$, $m(r\to\infty)\to 0$,
$S(0)=0$, and $S(r\to\infty)\to 1$.
Note that requiring $\bf m$ to lie in a fixed direction
does not restrict the 
minimization problem, since spatial variations in direction of
${\bf m}$ will increase the free energy.
With this ansatz, the energy takes the form:

\bea \label{3}
{\cal E}&=&\pi \int_0^\infty  \{
    (f'(r))^2 + \left( S'(r)\over r \right)^2  \nonumber \\
&&{}+ {(1-S)^2 f^2\over r^2}
        + (m'(r))^2 + \kappa^2 g m^2   \nonumber \\
&&{}+ \kappa^2 (1-f^2-m^2)^2  \}\, r\, dr.
\eea

$f,S,m$ satisfy the following Euler-Lagrange equations:
\bea
\label{Eq1}
 &&  -f''(r)-{1\over r}f'(r) + {(1-S(r))^2\over r^2}f \nonumber \\
&&\qquad \qquad- \kappa^2(1-f^2-m^2)f=0, \\
\label{Eq2}
&&   -S''(r) + {1\over r} S'(r) - (1-S)f^2=0, \\
\label{Eq3}
&&   -m''(r)-{1\over r} m'(r) + g\kappa^2 m \nonumber \\
&&\qquad \qquad-\kappa^2(1-f^2-m^2)m=0.
\eea

\begin{center}
\begin{figure*} 
\includegraphics[width=3.4in]{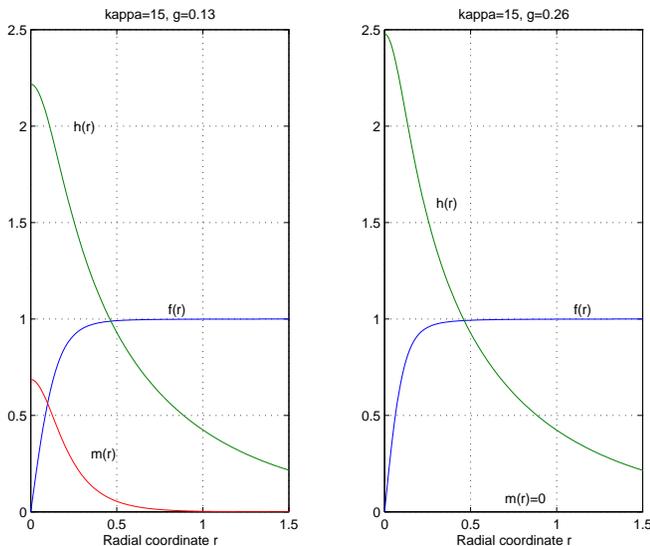}
\caption{Typical profiles for $f,h,m$ at $\kappa=15$.  On the left, 
the core is AF, while the core is normal on the right.}
\label{fig1}
\end{figure*}
\end{center}

By linearization at $r=0$  we obtain the standard
properties for vortex solutions: \cite{Berger-Chen,Plohr}
$f(r)\sim \alpha r$ and $S(r)\sim \beta r^2$ near $r=0$. 
Unconstrained minimization also yields $m'(0)=0$.  Using
Theorem 7.2 in Jaffe--Taubes \cite{Jaffe-Taubes} we see that
$f,S,m$ approach their asymptotic limits $r\to \infty$ exponentially.
In addition,  $f(r)$ and $S(r)$  are
monotonically increasing, while the $z$-component of the local field
$h(r)=S'(r)/r$  is monotonically decreasing.
The AF order parameter $m(r)$ is either positive and monotonically
decreasing or identically zero. \cite{ABG} 
These properties are illustrated in Figure~\ref{fig1}.

The results shown in Figure~\ref{fig1} and elsewhere in this paper come from numerical solutions of (\ref{Eq1})--(\ref{Eq3}). Our technique is to find free energy minimizers by
computing a negative gradient flow for a discretized free energy.
We approximate the improper integral in $\cal E$ by an integral over
a  finite
interval, large compared to the fundamental scales ($\lambda=1$
and $\xi=\kappa^{-1}$) in the problem.
We set a fixed boundary condition for $f,S,m$ at the right endpoint,
equal to their asymptotic values.
We then use finite elements to discretize the integral, with a fine mesh
concentrated at the vortex core.  For example, in Figure~\ref{fig3} we
divide the $r$-interval $[0,10]$ by 350 grid points, with 300 points
lying in $[0,2]$.  We choose standard linear elements, except
in the first interval near 0, where the behavior of $S(r)$ is
known to be quadratic.

We then explicitly calculate the gradient of the discretized
energy with respect to the coefficients $(f_k),(S_k),(m_k)$
of the base elements.  To find the minimal configuration,
we start with some  initial data, and flow
the coefficient vector $(f_k),(S_k),(m_k)$ against the
gradient until an equilibrium is found.
To solve the gradient flow numerically, we use a code developed by
Dr. N. Carlson of Los Alamos National Laboratory, which he
very kindly made available to us.  In this code the ODE is solved
by an implicit method, using second order backwards differences.
At each time step the resulting nonlinear system is solved by
a modified Newton method, and the time step is chosen based on
estimates of the local truncation error.

\smallskip

An important difference between our results and those of Arovas {\it et al.} is that our solutions have superspin magnitude strictly less than one in the core which allows the transition to be continuous.
To see this, let $u=f^2+m^2$, so $u$ solves the ODE
\bea\label{uEqn}
&&u'' + {1\over r}u' + 2\kappa^2(1-u)u= \nonumber\\
            && \qquad \qquad 2{(1-S)^2\over r^2}f^2 + 2g\kappa^2 m^2 + (f')^2 + (m')^2.
\eea
First, assume that the maximum value of $u$ is achieved at
$r_0\ge 0$.  Note that $u'(r_0)=0$, which follows from
standard calculus if $r_0>0$, and $u'(0)\le 0$ if $r_0=0$.
In addition, $u''(r_0)\le 0$
and the right-hand side of (\ref{uEqn}) must be strictly positive at $r_0$, so
$2\kappa^2 (1-u)u > 0.$  Since $u\ge 0$ by its definition,
we must have $u=f^2+m^2 <1$.  In the core region, we can say
even more.  Since $m$ is positive and decreasing, it attains
its maximum at $r=0$, and therefore  (\ref{Eq3}) implies
$$   
 0\ge  m''(0) = \frac{\kappa^2}2 (g-1+m(0)^2)m(0). 
 $$
When $m(0)>0$ we have the upper bound on the AF order parameter
at the core:
\be\label{mg}   m(r)^2\le m(0)^2 \le 1-g . \ee
Since $f(r)\sim r$ for $r\sim 0$, we conclude that the 
superspin magnitude is bounded by $\sqrt{1-g}$ at $r=0$.

\subsection{Phase transition}
Conventional superconducting vortices with
normal cores are obtained by minimizing 
 ${\cal E}$ under the constraint $m(r)=0$.
Note that when $m=0$ equations (\ref{Eq1})--(\ref{Eq3})
reduce to the well-known vortex equations of Abrikosov
which have been
extensively studied in the physical and
mathematical literature.\cite{Berger-Chen,Plohr,ABG,Abrikosov,Hu}. 
 We denote by $f_{\kappa,0},S_{\kappa,0}$ these normal-core
vortex solutions.

We find that there is a critical value $g^*_\kappa$ with $0.2<g_\kappa^*<0.3$
at which the stability of the normal-core solutions 
$f_{\kappa,0}$, $S_{\kappa,0}$ (with $m=0$) changes:
normal cores are stable (i.e., energy minimizing)
 when $g>g_\kappa^*$ and unstable for $0<g<g^*_\kappa$.
To see this, we linearize (\ref{Eq3}) at the normal core solution
to obtain an eigenvalue problem:
\be\label{linear}
 {1\over \kappa^2}\tilde m''(r)+{1\over \kappa^2 r} \tilde m'(r) 
+ (1-f_{\kappa,0}^2)\tilde m
      = g \tilde m.
\ee
Define $g^*_\kappa$ to be the ground-state eigenvalue, with
associated eigenstate $\tilde m_\kappa$.  From  Rayleigh's
principle, $\tilde m_\kappa$ is the minimizer of
\be\label{4}
-g_\kappa^* =  \min_{w\neq 0}{  \int_0^\infty  \left[
           \kappa^{-2} (w')^2   -(1-f_{\kappa,0}^2) w^2\right]\, r\, dr   \over
       \int_0^\infty   w^2\, r\, dr},
\ee
When $g$ crosses below the value $g^*_\kappa$,
the linearized operator associated with (\ref{Eq3}) possesses a negative eigenvalue,
and  the normal core solutions become unstable. 
In fact, by  perturbing the normal core solution $m=0$ by
a small multiple $\epsilon \tilde m$ of the eigenstate,
we see that  the energy is decreased 
by allowing a nonzero AF component when $g<g^*_\kappa$:
\bea\label{second var}
&&{\cal E}[f_{\kappa,0},S_{\kappa,0},\epsilon \tilde m_{\kappa}]= {\cal E}[f_{\kappa,0},S_{\kappa,0},0] + \nonumber \\
 && \qquad\qquad  
       {\epsilon^2\over 2}  (g-g^*_\kappa)\kappa^2\int_0^\infty 
               \tilde m_\kappa^2\, r\, dr   + O(\epsilon^3).
\eea

\begin{center}
\begin{figure*} 
\includegraphics[width=3.4in]{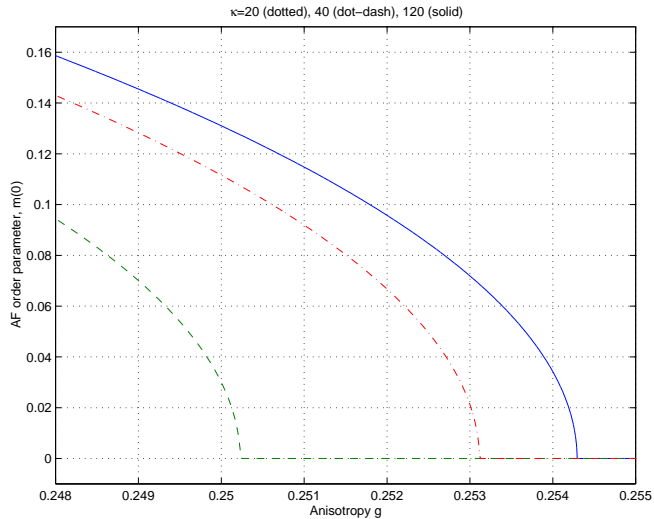}
\caption{Nucleation of antiferromagnetic cores for various $\kappa$}
\label{fig2}
\end{figure*}
\end{center}

\begin{center}
\begin{figure*}
\includegraphics[width=3.4in]{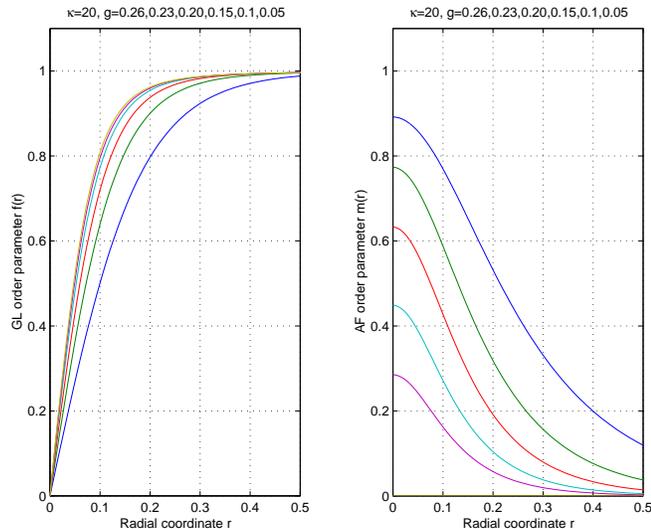}
\caption{$f,m$ for different values of $g$. As $g$ decreases, we see that the
radius of the vortex increases and that $m$ increases.}
\label{fig3}
\end{figure*}
\end{center}

Our numerical computations show that the transition is
continuous (second order) at $g^*_\kappa$.  Indeed, in Figure~\ref{fig2}
we plot a bifurcation curve, $g$ vs.~ $m(0)$ obtained by
numerical simulation at several values of $\kappa$.
Since $m(r)$ is a decreasing, positive function, the value 
$m(0)$ represents the maximum value of $m$.
Our analytic and numerical results predict that the
superspin parameter will lift continuously out of
the superconducting plane as $g$ is decreased through
the critical value $g^*_\kappa$, with its maximum value
approaching 1 as $g\to 0$ from above.  In Figure~\ref{fig3}
we illustrate the continuous growth of the AF component
with decreasing $g$.  We see also that the radius of the vortex core increases as anisotropy $g$ decreases. 

In computing the bifurcation curves shown in Figure~\ref{fig2} 
we allow $g$ to change
very slowly in time, $g(t)\sim 10^{-8}\, t$.  In this way, the
minimum point of the free energy moves, but very slowly compared
to the relaxation time of the gradient flow.  We ran the program with
many different values of $\kappa$ and grid spacings and always
observed the same qualitative behavior for the transition.  The
results obtained with slowly varying $g$ were checked against
solutions calculated at some fixed $g$ and found to be virtually
identical.
As a further check on our routines we made an independent
computation of the bifurcation point $g^*_\kappa$
using the Rayleigh formula (\ref{4}).  Using a finite element
discretization of (\ref{4}) and a standard eigenvalue solver
from EISPACK, we obtain good agreement with the observed
transition values in Figure~\ref{fig2}.  Some values are given
in Table \ref{gtable}.

\begin{table}
\begin{tabular}{|r|r|} \hline
$\kappa$     &     $g^*_\kappa$       \\ 
  \hline \hline
20       &     0.2502    \\ \hline
40      &  0.2531      \\ \hline
80   &  0.2541      \\  \hline
120   &  0.2543      \\ \hline
$\infty$   &  0.2545     \\  \hline
\end{tabular}
\label{gtable}
\caption{Values of $g^*_\kappa$. Computations were made with 1501 grid points.}
\end{table}

\begin{center}
\begin{figure*} 
\includegraphics[width=3.4in]{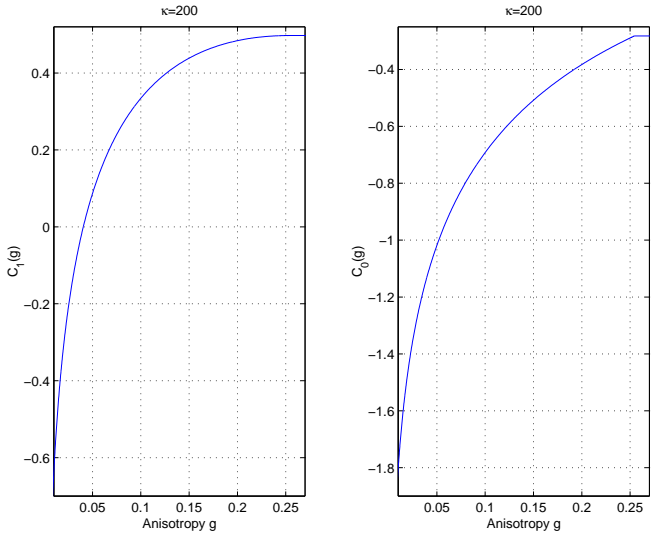}
\caption{Dependance of $C_1$ and $C_0$ on $g$}
\label{fig4}
\end{figure*}
\end{center}

Next we consider the dependence 
on  $g$ of the dimensionless vortex-line energy $\epsilon_1$ of an
AF core for large $\kappa$ ($\kappa =200$). 
A standard estimate \cite{Tinkham} shows that for large $\kappa$
\be \epsilon_1={1\over{4\pi}}\cal E
\mbox{$\sim {1\over 4}(\ln \kappa + C_1(g))$}.
\ee
Our numerics predict a value $C_1(g)=0.49739$ when
 $g>g^*_k$ (normal cores), which is in
good agreement with the value $C_1=0.4976$ obtained by
 Hu\cite{Hu}.  For AF cores, we observe that $C_1(g)$ decreases
as $g$ decreases ({\it cf.}~Figure~\ref{fig4}).

Similarly,  for large $\kappa$
\be \mbox{$h(0)\sim \ln\kappa + C_0(g)$}
\ee
with $C_0(g)=-.28192$ for $ g>g^*_k$ (in good agreement with Hu
\cite{Hu} who obtained $C_0=-.2823$) and with $C_0(g)$ decreasing as $g$
decreases ({\it cf.}~Figure~\ref{fig4}). So the values of $\epsilon_1$ and $h(0)$ decrease as the cores becomes more AF. 
Recall that the size of the core also increases as can be seen in Figure~\ref{fig3}. These results show that both $H_{c_1}$ and $H_{c_2}$ decrease as the core becomes larger and more antiferromagnetic.

\section{The extreme Type II limit}

To compare our results more directly with the work of Arovas et al \cite{Arovas}
we consider the extreme type II model.  In fact, for this simpler model
we may obtain rigorous analytical results concerning the transition
to AF cores. \cite{ABG} 
After rescaling by the correlation length $\xi=\kappa^{-1}$,
 the order parameters converge
(uniformly) to a fixed profile,
 $f_\kappa(\frac r \kappa)\to f_\infty(r)$ and $m_\kappa(\frac r \kappa)\to m_\infty(r)$ 
as $\kappa\to\infty$.  (Note that in this scaling $S_\kappa(\frac r \kappa)\to 0$.)
Passing to the limit in (\ref{Eq1})--(\ref{Eq3}) we obtain
 the extreme Type II equations for $f_\infty$ and $m_\infty$,
\bea\label{highk}
&&  -f''_\infty-{1\over r}f'_\infty + {1\over r^2} f_\infty 
   -(1-f_\infty^2-m_\infty^2)f_\infty =0, \\
&& 
\label{highk2}
 -m''_\infty-{1\over r}m'_\infty + g m_\infty 
   -(1-f_\infty^2 -m_\infty^2)m_\infty =0 .
\eea
For this model, it is rigorously proven \cite{ABG} that a continuous phase
transition to AF cores occurs at a critical value $g=g_\infty^*$,
with $g_\infty^*$ the ground-state eigenvalue of the linearized
equation (\ref{highk2}) at the normal core solution,
\be\label{5}
-g_\infty^* =  \min{  \int_0^\infty  \left[ (w')^2
             -(1-f_\infty^2) w^2  \right]\, r\, dr   \over
       \int_0^\infty   w^2\, r\, dr} 
\ee
At each value of $g$ there is exactly one stable vortex solution.
That stable profile has a normal core when $g\ge g_\infty^*$, and an AF core
when $g<g_\infty^*$.   In other words, the bifurcation diagrams
obtained by numerical simulation of (\ref{Eq1})--(\ref{Eq3}) with
$\kappa$ finite are proven valid for the limiting problem (\ref{highk}),
(\ref{highk2}).
This provides another check on our simulations.  For
$\kappa$ large we should see qualitatively similar behavior
between this model and that of (\ref{highk}), (\ref{highk2}).  Numerical computation of $g_\infty^*$ using the Rayleigh formula (\ref{5}) is consistent with the other calculated values of $g_\kappa^*$. (See Table \ref{gtable}.)
It also shows that vortices with AF cores are stable in a much broader
range than was found in Arovas {\it et al.} \cite{Arovas},  where $g^*_\infty$ is estimated as $0.0941$. This
discrepancy is due to the fact that the superspin magnitude is in fact far from $1$ in the core region (see
our analysis in Section 2.1), while the variational solution of  Arovas et al \cite{Arovas} assumes that AF
core solutions have superspins constrained to the unit sphere. \cite{footnote} 

\section{Phase Diagram}

\begin{center}
\begin{figure*} 
\includegraphics[width=3.4in]{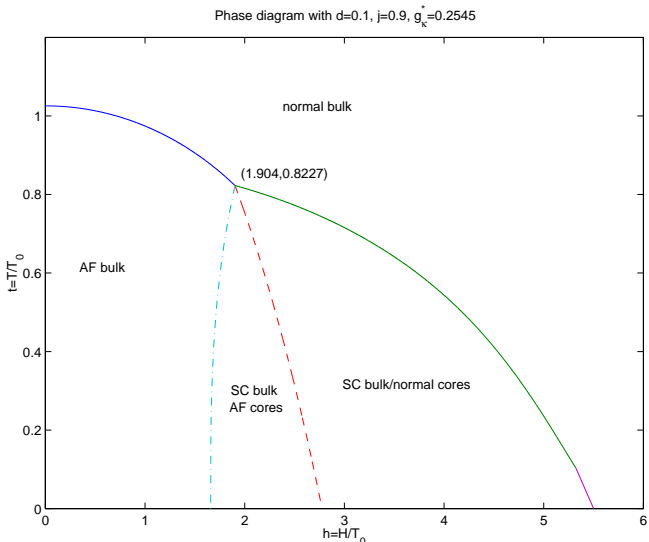}
\caption{Mean field phase diagram for the ``spin-flop'' model defined by Eq.~(\ref{mfh}).  The 
solid lines separating the AF and SC bulk phases from the high temperature disordered state represent continuous transitions.  The dot-dash line separating
AF bulk from ``SC bulk AF cores'' is first order, and the dashed boundary between ``AF cores'' and ``normal cores'' represents the continuous transition
obtained in this paper.}
\label{fig5}
\end{figure*}
\end{center}

The results of the previous two sections for $g^*_\kappa$ define a region in the phase diagram 
in which vortices with antiferromagnetic cores are stable.  To illustrate this statement, we add the (dashed) line for the stability boundary of vortices
with antiferromagnetic cores to the SO(5) mean-field phase diagram, assuming the relation (\ref{g}) between $g$ and $\mu$. The result is shown in
Figure~\ref{fig5}. 

The SO(5) mean-field phase diagram of Figure~\ref{fig5} is actually the mean field ``spin-flop" 
phase diagram for the three-component antiferromagnet with uniaxial anisotropy in a parallel magnetic field. In this analog system\cite{Zhang}, the
antiferromagnetic state with its Neel vector parallel to the field corresponds to the SO(5) AF state, while the ``flopped" antiferromagnetic state, with Neel
vector perpendicular to the applied field, corresponds to the SC state.  The field $H$ plays the role of the SO(5) chemical potential $\mu$.  The spin
Hamiltonian for this system is:
\bea\label{mfh}
&&{\cal H}=J\Sigma_{(m,n)} \left[ S_m^zS_n^z +j(S_m^xS_n^x+S_m^yS_n^y)\right] \nonumber \\
&&{}- D\Sigma_m (S_m^z)^2 -H\Sigma_m S_m^z.
\eea 
Where $d=D/J$ and $j$ are parameters which control the anisotropy. Figure~\ref{fig5} shows the 
mean field phase diagram for this model for specific values of these parameters (d=0.1, j=0.9). The unit of energy in this diagram is $T_0=zJ/3$ where $z$ is
the coordination number. The region labeled ``SC bulk, AF cores'' is bounded on the left by the first order SC-to-AF transition and on the right by the
stability boundary for AF cores.

At the mean field level one might expect to see antiferromagnetic cores throughout this region, 
with the amplitude of the antiferromagnetism being largest close to the first order boundary. However, as Bruus {\it et al}\cite{bruus} discuss, the spins in
the cores are likely to be only weakly correlated from one vortex to the next, and hence the signatures of antiferromagnetic vortex cores may be rather
subtle.  Bruus {\it et al} suggest that excitations of the cores could be studied by inelastic neutron scattering, while Arovas {\it et al}\cite{Arovas}
proposed studying the effect of antiferromagnetic cores on the $\mu$SR spectrum.

\section{Acknowledgements}

This work was supported by the Natural Sciences and Engineering Research Council of Canada. 
We thank the Brockhouse Institute
for Materials Research for supporting a workshop which brought
together physicists and mathematicians to discuss 
issues in superconductivity. 
AJB acknowledges the hospitality of the Aspen Center for Physics, where part of this paper was 
written, and support from the Superconductivity Program of the
Canadian Institute for Advanced Research.

\end{document}